# PRECISION QUANTUM HALL RESISTANCE MEASUREMENT ON EPITAXIAL GRAPHENE DEVICE IN LOW MAGNETIC FIELD

A. Satrapinski[1 a)], S. Novikov[2] and N. Lebedeva[2]

[1]Centre for Metrology and Accreditation, MIKES, P.O.Box 9, FI-02151 Espoo, Finland
[2]Micro and Nanoscience Laboratory, Aalto University, Micronova, Tietotie 3, Espoo, Finland

**ABSTRACT.** Precision quantum Hall resistance (QHR) measurements were performed on large-area epitaxial graphene device at low magnetic fields ($B$ = 2 T - 8 T) at temperature $T$ = 1.5 K. Hall resistance was measured using Cryogenic Current Comparator resistance bridge with high biasing current $I_{sd}$ = 40 μA. The results showed that at $B$ = 8 T the relative deviation of Hall resistance from the expected quantized value $h/2e^2$ is within experimental uncertainty of 3.5 parts in $10^8$ and remained below 0.35 parts per million (ppm) down to $B$ = 3 T.

Application of unique electronic properties in graphene (Ref. 1,2,3) and ultra thin carbon layers is very perspective for future electronic devices and for electrical metrology. One of the main physical properties of graphene is anomalous "half-integer" Quantum Hall Effect (QHE), where Hall resistance is quantized on plateaus $R_H$ = ± ($R_K/4$)/($n$+0.5), where $R_K$ = $h/e^2$ is von Klitzing constant, $h$ is Plank constant, $e$ is elementary charge, and $n \geq 0$ is an integer. The evident application of graphene is the QHE resistance standard (Ref. 4 - 9), which is based on quantized values of Hall resistance conventionally indexed as $R_H(i) = h/(e^2 i)$ where $i$ is an integer filling factor. Recently a big progress has been achieved in precision measurement of QHE in two-dimensional electron gas (2DEG) in graphene. An accuracy of 3 parts in $10^9$ (ppb) for the resistance quantisation in SiC-based QHE device (Ref. 4) on $R_H(2) = h/(2e^2) \approx 12.9$ kΩ plateau at magnetic field $B$=14 T, biasing current $I_{sd}$=12 μA, and temperature $T$=0.3 K has been demonstrated for 160 μm x 35 μm sample in indirect measurement. In a direct comparison between GaAs and epitaxial SiC graphene (in high magnetic fields), the equivalence of the quantized Hall resistance (QHR) has been demonstrated with a relative uncertainty of 8.6 parts in $10^{11}$ and the invariance of $R_H$ (2) on materials has been confirmed (Ref. 8). These results affirmed and proved the use of epitaxial SiC graphene as a future quantum resistance standard.

In (Ref. 4) it was shown that in a QHR device based on graphene on SiC, covered with bilayer polymer and illuminated with UV light, the carrier density can be reduced down to $2 \cdot 10^{10}$ cm$^{-2}$ while low-temperature mobility is increased up to 16000 cm$^2$/Vs. After illumination of the sample, $R_H(2)$ plateau extended even down to magnetic field of 1 T at $T$=0.3 K. However, no precision measurements of quantized Hall resistance in graphene in low magnetic fields below about 10 T have been carried out before our work.

In this work we have investigated SiC graphene samples with a larger area (800 μm x 200 μm) than in earlier studies. Our aim was to investigate the possibilities of precision Hall measurement with higher measurement currents and to study the quantisation of 2DEG in SiC graphene Hall device in lower magnetic field and at higher temperatures.

We have performed accurate Hall resistance measurement at magnetic fields between 2 T and 9 T. Hall plateau was observed even at the lowest field, 2 T, at temperature as high as 1.5 K. Accurate measurements have been carried out with our precision ac Cryogenic Current Comparator (CCC) resistance bridge (Ref. 10) with relatively high biasing current, $I_{sd} = 41$ μA.

Graphene film was grown by "GraphenSic AB" company on Si-face of 4H-SiC substrate (Ref. 11,12). AFM topography of the graphene film showed existence of large flat terraces on the surface. The height of these terraces was around 1 nm. Thickness of the film was estimated from Raman spectroscopy, which confirmed the presence of a single layer of graphene and the low defect density in graphene film.

Patterns for the Hall bars and the contacts were made using laser photolithography over AZ5214 resist. Reactive ion etching in argon-oxygen plasma was used to remove the graphene layer from uncoated areas. Totally 18 devices with different dimensions were fabricated on one chip. The device with a large area (800 μm x 200 μm) was chosen for the QHR measurements. Configuration and numeration of the contacts for the device are presented in FIG. 1 (a). View of the chip assembled on TO-8 header is presented in FIG. 1 (b). The direction of the current channel in Hall bar device on the chip was chosen according to AFM image with the current channel directed along the terraces.

After fabrication of the Hall bars and contacts, the chip with the Hall devices was covered by two polymers, first by 300 nm of PMMA resist and second by 300 nm of ZEP520A resist. These resist layers were used for photochemical gating of graphene (Ref. 13) and for control

a)e-mail: Alexandre.satrapinski@mikes.fi

of the carrier concentration in graphene by illumination with UV light. Longitudinal resistance, $R_{xx}$, of the graphene sample was controlled and adjusted at 293 K using UV light from a deuterium broad band lamp with wavelength 240 nm and optical power 0.16 mW. Graphene sample was illuminated during several runs, with short exposure time (20 s – 60 s) until no change of resistance was observed. Total square resistance was increased by a factor of about two (at $T$=293 K), indicating the reduction of carrier concentration.

Magneto-transport measurements were performed both before and after illumination by UV light. An 11 T superconducting magnet system with variable temperature insert and with temperature control down to 1.5 K was used for the measurements. The sample was cooled down to 1.5 K slowly, during two days. Four-terminal resistance measurements of the Hall resistance $R_{xy}$ and longitudinal resistance $R_{xx}$ on different contact pairs were carried out with precision multimeters using dc current $I_{sd}$=10 µA. The results of the $R_{xy}$ and $R_{xx}$ measurements before (dashed line) and after (solid line) illumination with UV light are presented in FIG. 2 (a) and (b). $R_{xy}$ measured using contacts 13h – 13l (see FIG. 1 (a)) is presented in FIG. 2 (a). The device exhibits a well-defined plateau at $R_{xy} \approx h/2e^2$. It is seen in FIG. 2 (a) that for non-irradiated sample (higher carrier concentration), the plateau extends down to about 5 T, and for irradiated sample (lower carrier concentration) it extends down to about 1.8 T.

Solid line in FIG. 2 (b), in positive magnetic field direction, presents the results of the measurements carried out on 200 µm x 200 µm area (contacts 11l - 10l) and, for the negative magnetic field direction, on the area 400 µm x 200 µm (contact pairs 11l-13l). The values of $R_{xx}$ on the negative field are scaled to present the square resistance. Shubnikov-de Haas oscillations in $R_{xx}$ corresponding to $h/6e^2$ and $h/2e^2$ plateaus of $R_{xy}$ are clearly seen for non-illuminated sample in FIG. 2 (b). Weak localisation peak in $R_{xx}$ at zero magnetic field is observed for both low and high carrier concentrations, but the peak is sharper for low carrier density. Nonsymmetry in $R_{xx}$ for positive and negative magnetic field directions can be explained by formation of electron-hole puddles, changing the effective conducting area (Ref. 14) and possibly by imbalanced carrier injection from the graphene electrodes caused by misalignment of the electrode and channel neutrality points (Ref. 15).

In magnetic fields corresponding to Hall plateaus, the longitudinal resistance falls close to zero. To check a dissipationless state of a 2DEG on a plateau for the sample with high carrier concentration and to define the breakdown current, the measurement of $R_{xx}$ versus current were performed on contact pairs 11h–13h and 11l - 10l with the applied current in the range

[a)]e-mail: Alexandre.satrapinski@mikes.fi                                    3

10 μA – 200 μA. For currents below 150 μA the measured value of $R_{xx}$ (at $B$ = -8.0 T, and at $T$=1.42 K) was less than 1 mΩ. Measurement of $R_{xx}$ for the sample with low carrier concentration (after illumination) was performed only for currents $I_{sd}$ up to 50 μA and for $B$ = - 8 T and $B$ = 2.5 T. It was found that $R_{xx}$ was less than 10 mΩ for $B$ = -8 T and less than 230 mΩ for $B$ = 2.5 T.

FIG. 3 the mobility versus carrier density, estimated from the Hall measurements in low field and $R_{xx}$ measurement at 1.5 K, before and after UV illumination is presented. Before illumination with UV light, the carrier concentration was estimated as $4.5 \cdot 10^{11}$ cm$^{-2}$ and the mobility values measured using different contacts varied between 6850 cm$^2$/Vs and 8560 cm$^2$/Vs. After illumination, the carrier concentration was reduced to $6 \cdot 10^{10}$ cm$^{-2}$ and the square resistance measured using different contact pairs varied between 3.67 kΩ and 6.98 kΩ, which corresponds to a significant variation in mobility, from 16050 cm$^2$/Vs to 30520 cm$^2$/Vs. The reason for such a big scatter can be non-equilibrium distribution of carrier density along the channel and inhomogeneity of the graphene film due to residual defects and impurities.

Contact resistances were estimated from three-terminal resistance measurements performed on $h/2e^2$ plateau, at $B$ = -6.5 T, $T$ = 1.5 K. It was found that the contact resistance in current contacts is of the order of 20 Ω.

Precision measurements at $R_H(2)$ plateau were performed using CCC based ac resistance bridge (Ref.10). Relative deviation of $R_H(2)$ in graphene from the expected value $R_K/2$, d$r_H$ = $(R_H(2) - R_K/2)/(R_K/2)$, was determined. In this measurement indirect comparison of the $R_H(2)$ value in graphene to the realization of $R_H(2)$ in GaAs device was done using a stable 100 Ω standard resistor that had been calibrated against $R_H(2)$ in a GaAs device. Combined expanded uncertainty of evaluated value at 95 % confidence level is estimated as 2.8 parts in $10^8$.

In FIG. 4 the deviation d$r_H$ measured using 13l-13h Hall contact pair at $T$=1.5 K, $I_{sd}$ = 41 μA and in magnetic field range 2.5 T – 8 T, of the illuminated (open triangles) and non-illuminated sample (solid diamonds) is presented. At 8 T the direction of magnetic field was opposite to that used in lower field. For the illuminated sample, with low carrier density, the deviation at $R_H(2)$ from the expected value did not exceed 0.35 ppm in the range 3 T – 4 T, and d$r_H$ increased up to 1.9 ppm at magnetic field 2.5 T. The observed deviation for illuminated sample at $B$ = -8.0 T was about 0.15 ppm. The deviation is somewhat larger than the combined expanded uncertainty of the measurements which varied between 0.15 ppm and

a)e-mail: Alexandre.satrapinski@mikes.fi

0.30 ppm at 95 % confidence level. Statistical uncertainty was larger than normally due to bigger noise from graphene sample and short measurement time. The deviation could be possibly caused by the increased dissipations (i.e increased $R_{xx}$) and spatial fluctuations in the low carrier density sample, destroying QHE regime (Ref. 6). Exact reasons for the observed deviation will be studied in further research.

The results of precision measurement which were carried out at carrier density $4.5 \cdot 10^{11}$, before illumination with UV light, are shown in the inset of FIG.4. The measurement was carried out using 13l-13h Hall contact pair at $B$=-8 T, $T$=1.5 K and $I_{sd}$ = 41 μA. The value for d$r_H$ estimated from this measurement is +0.016 ppm with combined expanded uncertainty of 0.035 ppm (at 95 % confidence level). The measured value of d$r_H$ at other Hall contacts (10l-10h and 11l-11h) was larger but did not exceed 0.5 ppm. This deviation was caused probably by misalignments of the Hall bars along the channel, which were noticed as an increased (by about one order of magnitude) $R_{xy}$ on these Hall bars (10l-10h and 11l-11h) at $B$ = 0. The observed larger deviation for these contact pairs can be partly explained also by the influence of increased contact resistances at different contacts. Contact pair dependence also can indicate partial non-homogeneity of the carrier density in tested sample. Interestingly, the voltage noise on potential contact pairs of the graphene device was significantly larger than the noise level measured on the potential contacts of a conventional GaAs device at the $h/2e^2$ plateau. The increased noise from the sample was the main factor limiting statistical uncertainty of measurement.

In conclusion, a wide $h/2e^2$ plateau down to very low (2 T) magnetic field at temperatures around 1.5 K was observed in the Hall resistance of a large area epitaxial graphene device fabricated on SiC substrate. Measurement results showed that the quantization of Hall resistance in 2DEG at magnetic field as low as 2 T is possible due to application of photochemical gating and reduction of carrier concentration down to $6 \cdot 10^{10}$ cm$^{-2}$. Together with the reduction of carrier concentration, electron mobility increased up to 30520 cm$^2$/Vs, but the variation of mobility was larger than in the original state with higher carrier concentration. The deviation of quantized Hall resistance $R_H(2)$ from $R_K/2$, measured on the device with carrier concentration $5 \cdot 10^{11}$ cm$^{-2}$ at $T$ = 1.5 K with applied current $I_{sd}$= 41 μA and at $B$ = -8 T was (+0.016 ± 0.035) ppm. The value of $R_H(2)$ measured on the same device but using different Hall contact pairs varied but relative deviation from the expected value did not exceed 0.5 ppm. Precision QHR measurements were carried out for the first time at low magnetic fields (2.5 T – 4 T). The results on the device with reduced carrier density showed

[a)]e-mail: Alexandre.satrapinski@mikes.fi

that relative deviation of $R_H(2)$ from $R_K/2$ remained below 0.35 ppm for magnetic fields down to $B = 3$ T.

The measurement results demonstrate good prospects and perspectives of application of epitaxial graphene for new quantum resistance standard with relieved experimental conditions.

## ACKNOWLEDGMENTS

We would like to thank R. Yakimova, S. Kubatkin, A. Tzalenchuk, S. Lara-Avila and A. Manninen for assistance and useful advices.

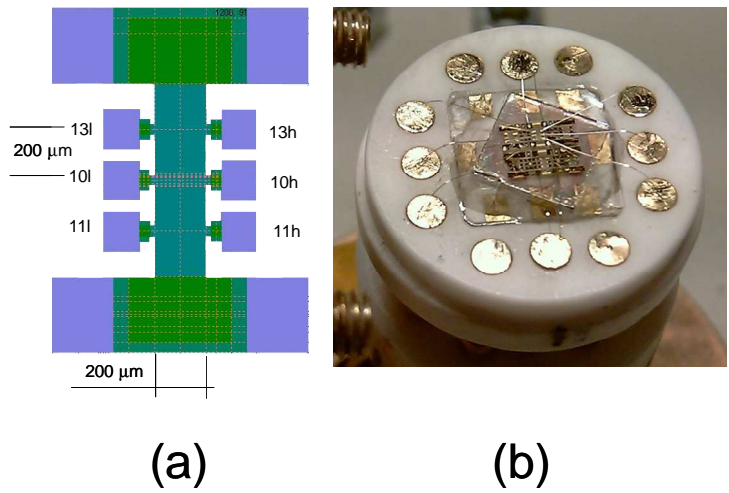

(a)     (b)

**FIG. 1. (a) Configuration and numeration of the contacts on one Hall device. (b) View of the chip with 18 Hall devices. One large area sample is bonded on TO-8 header.**

[a)]e-mail: Alexandre.satrapinski@mikes.fi

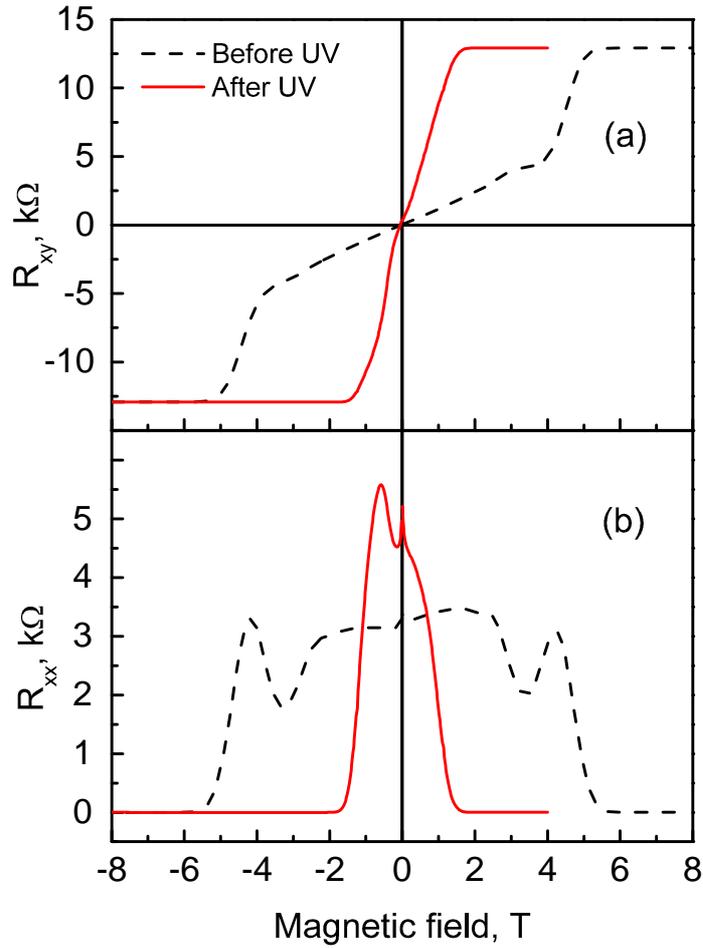

FIG. 2. (a) Hall resistance, $R_{xy}$, and (b) longitudinal square resistance, $R_{xx}$, versus magnetic field at $T = 1.5$ K and $I_{sd} = 10$ μA before (dashed line) and after (solid line) illumination of the sample with UV light. $R_{xy}$ was measured using contacts 13h-13l (see FIG. 1 (a)), and contacts 11l – 10l (11l – 13l) were used in $R_{xx}$ measurement at B>0 (B<0).

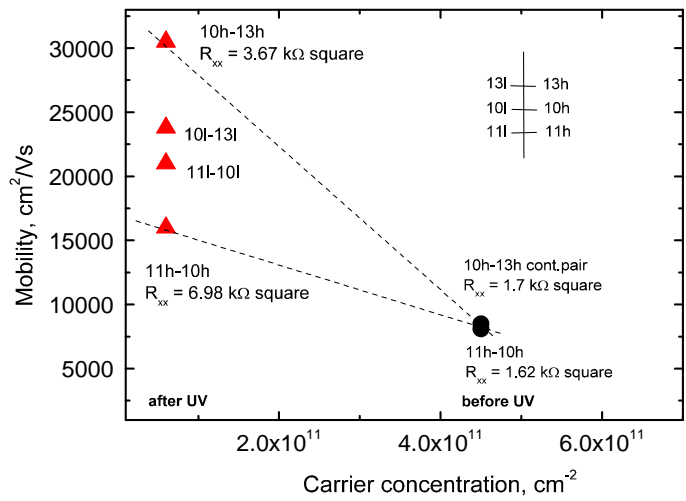

**FIG. 3. Mobility variations on different contact pairs estimated from resistivity and Hall measurement before (solid circles)) and after (solid triangles) illumination with UV light.**

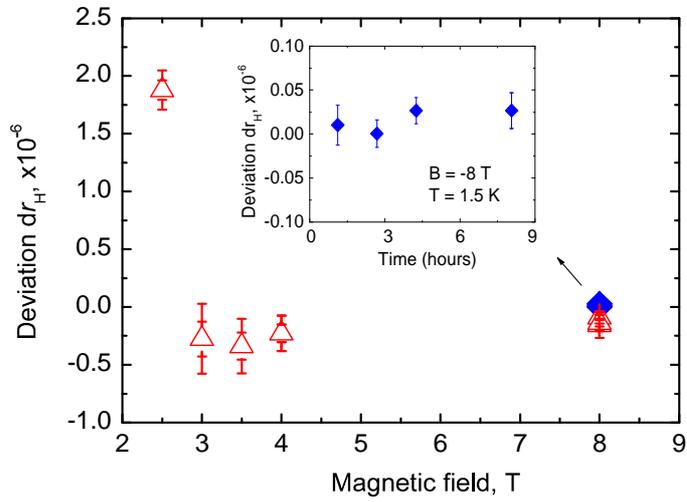

**FIG. 4. Deviation d$r_H$ measured in magnetic field range 2.5 T – 8 T at 13l-13h Hall contact pair after (open triangles) and before (solid diamonds) illumination with UV light. Horizontal axis is the absolute value of magnetic field. In inset the results of the deviation measurement of non-illuminated sample at the same Hall contacts at $B$=-8 T, $T$=1.5 K and $I_{sd}$ = 41 μA are presented.**